# Varifocal zoom imaging with large area focal length adjustable metalenses

**Authors:** Shane Colburn[1], Alan Zhan[2], Arka Majumdar[1,2]*

**Affiliations:**

[1]Department of Electrical Engineering, University of Washington, Seattle, Washington 98195, USA.

[2]Department of Physics, University of Washington, Seattle, Washington 98195, USA.

*Correspondence to: arka@uw.edu

**Abstract:** Varifocal lenses are essential components of dynamic optical systems with applications in photography, mixed reality, and microscopy. Metasurface optics has strong potential for creating tunable flat optics. Existing tunable metalenses, however, typically require microelectromechanical actuators, which cannot be scaled to large area devices, or rely on high voltages to stretch a flexible substrate and achieve a sufficient tuning range. Here, we build a 1 cm aperture varifocal metalens system at 1550 nm wavelength inspired by an Alvarez lens, fabricated using high-throughput stepper photolithography. We demonstrate a nonlinear change in focal length by minimally actuating two cubic phase metasurfaces laterally, with focusing efficiency as high as 57% and a wide focal length change of more than 6 cm (> 200%). We also test a lens design at visible wavelength and conduct varifocal zoom imaging with a demonstrated 4x zoom capability without any other optical elements in the imaging path.

**Main:**

Metasurfaces have fostered substantial interest in the optics and photonics communities in recent years. These ultrathin elements comprise arrays of subwavelength-spaced optical antennas that can apply spatially varying transfer functions on incident wavefronts[1–4], including those of lenses[5–11], holograms[12,13], polarization elements[14], vortex beam generators[14–19], and blazed gratings[20,21]. By only changing the lateral geometry and orientation of these optical antennas, the local transmission or reflection coefficient of a metasurface can be tailored for specific applications. Most demonstrated metasurface devices to date, however, have been static in nature. For metalenses, focal length tuning over a wide range is of substantial interest in photography, microscopy, mixed reality, and optical communications. Stretching of metalenses on flexible substrates[22–25] is one route to accomplish this, but this entails constant application of an external force to counteract the substrate's restoring force. Furthermore, electrical control of such systems requires high voltages (kV range) as the tuning mechanism relies on a capacitive electrostatic force to compress an elastomer[25]. Microelectromechanical systems (MEMS)-based tuning is promising, with recent results adjusting the angular orientation of a metalens[26] or demonstrating large changes in optical power by actuating a metalens axially in a compound lens system[27]. Unfortunately, MEMS devices entail extensive and challenging fabrication, and the actuation mechanisms cannot be scaled to the macroscale sizes necessary[28,29] for many applications requiring large apertures and focal lengths, such as imaging at high resolution or for eyeglasses and mixed reality displays[30,31]. With MEMS, the voltage required to induce sufficient displacement of large area metalenses would cause electrical breakdown and device failure.

Here, we develop a large area tunable focal length metalens system using an Alvarez lens design[32,33], combining two separate cubic metasurfaces that under lateral actuation give rise to a rapid and nonlinear change in focal length (Figure 1A). Unlike most previous metasurface works, we fabricate our device using high-throughput stepper photolithography, circumventing the scalability issues of electron-beam lithography to build a large area (1 $cm^2$ aperture) device. Our process flow relies on a custom-developed compression algorithm which can substantially reduce the complexity of layout files, enabling us to create a metalens with nearly 120 million scatterers with, to the best of our knowledge, the largest focal length range demonstrated to date. We use a versatile silicon nitride cylindrical nanopost platform, which is polarization-insensitive, and well suited for efficient operation from the visible to the infrared. In this work, our Alvarez metalens is actuated by hand using translation stages; however, electrical actuation is well within the means of commercially available miniature stepper motors[34]. With the wide aperture of our system, we demonstrate its utility for varifocal zoom imaging without requiring any additional elements (e.g., objectives or tube lenses) in the optical path, achieving a magnification range with 4x zoom capability in our experiments.

**Results:**

*Design:*

The phase profile of a spherical singlet metalens follows a quadratic form which is inversely proportional to its focal length. An Alvarez lens instead comprises two separate cubic phase plates that in conjunction give rise to a tunable focal length lens when the plates are actuated laterally with respect to one another. These phase plates are typically implemented using freeform glass surfaces or multi-level diffractive optics, entailing sophisticated and expensive fabrication. For our design, we use two flat metasurface phase plates where one plate obeys a cubic polynomial function and the other follows the same function but with opposite sign. When the plates are aligned along the optical axis and positioned parallel to one another, under lateral actuation the total phase profile imparted on an incident wavefront is a quadratic function with changing focus. The phase profiles of the regular and inverse metasurfaces are

$$\varphi_{reg}(x,y) = -\varphi_{inv}(x,y) = A\left(\frac{1}{3}x^3 + xy^2\right), (1)$$

where $A$ is a constant with units of inverse cubic length and $(x,y)$ represents the in-plane position. Translating these phase masks by a displacement $d$ in opposite directions, we obtain a quadratic form as below

$$\varphi_{Alvarez}(x,y) = \varphi_{reg}(x+d,y) + \varphi_{inv}(x-d,y) = 2Ad(x^2+y^2) + \frac{2}{3}d^3, (2)$$

Neglecting the constant $d^3$ phase term and relating the quadratic term to the phase of a spherical singlet, we find the focal length as a function of displacement as

$$f(d) = \frac{\pi}{2\lambda A d}, (3)$$

where $f$ is the focal length and $\lambda$ is the wavelength. The inverse dependence of the focal length on the displacement $d$ induces a nonlinear and rapid change in focal length for small displacements. For a design wavelength of 1550 nm, we select the value $A = 6.756 \times 10^9 m^{-3}$ such that by changing $d$ from 1 mm to 4 mm, we can tune the focal length continuously from 3.75 cm to 15 cm.

To implement our tunable lens, we need to design dielectric scatterers capable of supporting high efficiency operation while remaining compatible with the spatial resolution achievable with photolithography. For the designed focal length, a large aperture is required to cover the necessary actuation range and achieve a numerical aperture (NA) high enough to image with sufficient resolution. While large area metalenses with moderate to high NA do exist, these devices rely on expensive and time-consuming electron-beam lithography, precluding widespread commercial adoption. Recently, mass manufacturing-compatible large area metalenses at 1550 nm were reported[35]. Our process flow is similar to this work[35], however, we use a silicon nitride nanopost platform and demonstrate devices on both silicon and quartz substrates, enabling visible wavelength operation in addition to 1550 nm. To work within the constraints of our university cleanroom's stepper lithography system, we limited the minimum diameter of our silicon nitride cylindrical nanoposts (Figure 1B) to 500 nm and designed scatterers using rigorous coupled-wave analysis (RCWA)[36]. Figure 1E shows the simulated transmission coefficient at 1550 nm of our designed 2 μm thick silicon nitride nanoposts on a silicon substrate. We use a lattice spacing of 1.3 μm and have diameters ranging from 500 nm to 1.1 μm. As evidenced by the minimal variation in transmission coefficient over a wide range of lattice periodicities (Figure 1C-D), we can approximate our nanoposts as weakly coupled dielectric scatterers, justifying our subsequent use of the unit cell approximation in designing the metasurface[8]. These nanoposts can be understood as behaving like truncated circular waveguides in which the discontinuities in refractive index at the top and bottom interfaces of the scatterers produce Fresnel reflections and low quality factor Fabry-Perot resonances. Together these modes produce the nanoposts' complex transmission coefficient.

*High-throughput large area fabrication:*

To make our metasurface cubic phase plates, we used the calculated transmission coefficient data as a lookup table, mapping the desired phase to the corresponding nanopost diameter. Fabricating the desired metasurfaces required manufacturing a reticle in accordance with a layout file, such as a GDSII, detailing the positions and diameters of nearly 120 million nanoposts. Whereas layout files for typical metasurface designs usually contain individual cells for each nanopost due to the small number of individual posts, for the exceedingly large number of elements in our design, we had to develop an algorithm (see supplementary materials for details) based on hierarchical cell references to reduce the required memory. Minimizing the memory is critical, as layout files must undergo computationally demanding processing, such as fracturing, to convert the data into the proper format for manufacturing a reticle. With the number of elements increasing quadratically with a linear increase in aperture width, layout file compression is crucial to be able to support large area metalenses. By writing our layout file using our algorithm and converting to an OASIS file, we achieved more than a 2600x reduction in memory. While a previously developed metasurface layout file compression algorithm[35] showed an even larger reduction, our algorithm is more general in that it does not require any symmetry in the layout and can be used for general phase masks such as those for holograms or our cubic masks.

Figure 2A schematically summarizes our fabrication process flow, including deposition, spin coating, stepper lithography, hard mask patterning and etching, and mask removal. Figures 2B and 2C show a standard 100 mm wafer after our exposure step and an etched and cleaved cubic metasurface phase plate with a hand for scale respectively. In Figure 2D and 2E, we can see scanning electron micrographs of the fabricated nanoposts from normal and diagonal (45°) views respectively. In addition to the designed metasurface Alvarez lens, our reticle also included several static singlet and vortex beam-generating metalenses to demonstrate the versatility of our nanopost design and fabrication process. In characterizing these devices, we saw close to diffraction-limited performance and successful generation of vortex beams with different orbital angular momentum states (see supplementary Figure S1 for details).

*Focal length tuning and varifocal zoom imaging:*

We experimentally verified the tunable behavior of our Alvarez metalens system by laterally displacing the regular and inverse cubic metasurfaces with respect to one another (see supplementary Figure S2A for a schematic of the measurement setup). We displaced the metasurfaces over a 2.75 mm range, translating to a nonlinear change in focal length over a 6.62 cm range at 1550 nm, matching closely with the theoretical focal length (Figure 3A). With the same reticle layout, we also fabricated a lens on a quartz substrate to operate at 633 nm wavelength. While this visible regime device is not strictly a metasurface due to its super-wavelength lattice periodicity, the near-wavelength spacing still enables a wide range of phase shifts as a function of diameter, even with fixed nanopost thickness (Figure 3B). For this simulation, the nanoposts have the same lattice constant as before, but we use a lower thickness of 1.5 µm, which exhibited higher transmission amplitude. In having the same spatial arrangement of nanopost positions and diameters by using the same reticle as for the silicon substrate design, the metalens will still focus; however, as chromatic aberrations in metasurfaces are primarily a result of phase-wrapping disontinuities[37], in illuminating at 633 nm the phase function will exhibit discontinuities which will induce a chromatic focal shift. The resultant focal length of the metalens on quartz can be estimated via equation (3) and the strong agreement of this theoretical focal length and the experimentally measured focusing (see supplementary Figure S2B for a schematic of the measurement setup) confirms this behavior (Figure 3C). The visible lens design with its super-wavelength lattice constant does, however, come at the cost of producing higher diffraction orders that are absent for devices on a subwavelength lattice. The 1550 nm and 633 nm designs achieved focusing efficiencies of 57% and 15% respectively at a displacement $d$ of 2.5 mm. The large efficiency drop at 633 nm is attributed to light being lost to these additional diffraction orders, and when considering the focusing efficiency of the zeroth order beam alone (i.e., neglecting light lost to higher order diffraction, see Methods for details), we achieve an efficiency of 58%, close to that of the 1550 nm device. We note that our university cleanroom sets a lower limit on our achievable lattice constant, but subwavelength lattices for visible frequencies are well within the capabilities of state-of-the-art deep-UV lithography systems. With a subwavelength lattice constant, we could suppress these higher diffraction orders and increase the efficiency of the 633 nm design.

      Our widely focus-tunable lens is well suited for imaging with different values of magnification for varifocal zoom applications. To examine the imaging performance of our device, we illuminated a 1951 Air Force resolution test chart with a 625 nm LED in transmission and imaged the pattern directly onto a camera with our tunable lens on a quartz substrate without the use of supplemental optics (see supplementary Figure S3A for a schematic of the measurement setup). By fixing the test chart 30 cm away and tuning the focal length from 10 to

20 cm (1.8 mm actuation of each metasurface) and appropriately shifting the camera to the image plane, we provided magnifications ranging from 0.5x to 2x, achieving a 4x zoom range (Figure 4A). We repeated this measurement for imaging a Mona Lisa pattern prepared on standard printer paper by scattering the LED light off the pattern (see supplementary Figure S3B for a schematic of the measurement setup). To demonstrate the narrow actuation range required for changing the optical power of our metalens and its effect on imaging, we varied the degree of lateral misalignment of the two plates over a small range (-250 µm to + 250 µm) and recorded a video of the Air Force pattern with this actuation in real-time (see supplementary Video 1). Snapshots at specific levels of misalignment are also provided in Figure 4B. The narrow range required to actuate the device demonstrates the sensitivity of this tuning method, where the nonlinear change in focal length is very abrupt as a function of displacement.

**Discussion:**

Our tunable metalens system demonstrates a large change in optical power (20.8 diopters at 1550 nm and 9.2 diopters at 633 nm) with, to the best of our knowledge, the largest focal length range (6.62 cm at 1550 nm and 32.4 cm at 633 nm, 205% and 378% changes, $\frac{f_{max}-f_{min}}{f_{min}}$, respectively) for an optical metasurface demonstrated to date. This large tuning range is enabled not only by the nonlinear relationship between focal length and displacement for Alvarez lenses, but also by our developed stepper lithography platform and silicon nitride nanoposts, which can provide elements with much wider apertures using methods compatible with mass manufacturing. Currently, however, the metasurfaces in our demonstrated device are actuated by hand using micrometer translation stages. Such stages would be incompatible for any portable lens platform. Whereas the wide aperture of our device is one of its primary benefits, the corresponding increased mass of our optical element precludes MEMS-based actuation as demonstrated with other tunable metasurface systems[26,27]. The actuation is, however, well within the capabilities of commercial off-the-shelf stepper motors[34], such as those used to drive small masses or gears in wristwatches. Integration of our large area metasurface cubic phase plates with these actuators would provide rapid and low-power (zero static power dissipation) focal length-tunable metalenses.

The developed tunable lens also demonstrated varifocal zoom imaging, adjusting magnification from 0.5x to 2x, with large (10's of centimeters) object and image distances. While we did not demonstrate a true parfocal zoom with our device, integration of two such Alvarez lenses[38] would allow for zoom imaging with stationary optical components and fixed object and image positions. Integrating these two separate devices in a compact form factor would require modification of the tunable focal length range and therefore increasing the constant $A$ in equation (3) to provide a design with shorter focal lengths.

The reported system demonstrates metalenses with a wide focal length tuning range and varifocal zoom imaging capability requiring minimal lateral actuation. Expanding on our previous work integrating two cubic metasurfaces[33], this system provides a tunable metalens with nearly 120 million nanoposts, more than 1300 times the number in our previous work, attributable to the stepper photolithography-compatible processing we developed with a versatile nanopost platform. This wide tuning range and varifocal zoom capability could find applications in microscopy, planar cameras, mixed reality, and light detection and ranging (LIDAR). The demonstrated metalens provides a pathway for metasurfaces to become a viable commercial technology, leveraging existing mass manufacturing processes and commercial off-the-shelf electronics to reduce the mass and volume of optical systems while retaining sufficient imaging quality and providing a low-power tuning mechanism.

**Methods:**

Simulation and Design:

The silicon nitride nanoposts were simulated using the Stanford S4 rigorous coupled-wave analysis (RCWA) package[36]. The validity of the unit cell approximation was evaluated by calculating the transmission amplitude and phase as a function of both diameter and lattice constant. For our 1550 nm simulation, we set the refractive indices of silicon and silicon nitride to 3.476 and 1.996 respectively. For our 633 nm simulation, we set the refractive indices of silicon dioxide and silicon nitride to 1.457 and 2.039 respectively. Subsequent metasurface designs were evaluated as complex amplitude masks based on the RCWA data and simulated by evaluating the Rayleigh-Sommerfeld diffraction integral by means of an angular spectrum propagator.

Fabrication:

We used a dictionary of 6 fixed nanopost designs to make a hierarchy of cell references when generating our metasurfaces' layout file to substantially cut the required memory. Our process began with either a 100 mm silicon or quartz wafer, depending on whether the infrared or visible lens design was being fabricated. Our reticle was fabricated by Toppan Photomasks, Inc. and we transferred our pattern onto a film of AZ Mir 701 11 cps spun on top of a 2 μm PECVD silicon nitride layer using a 5x reduction stepper lithography system (Canon FPA-3000 i4). After development in AZ 300, we evaporated and then lifted off a 150 nm layer of aluminum to form a hard mask. We then etched the exposed nitride layer using a $CHF_3$ and $SF_6$ chemistry using an inductively-coupled plasma etcher system. We removed the remaining aluminum by etching in AD-10 photoresist developer. To capture the scanning electron micrographs, we sputtered an 8 nm Au/Pd film as a charge dissipation layer, which was subsequently removed using type TFA gold etchant.

Focal Length Measurement:

To measure the focal length of our 1550 nm metalens, we imaged the focal plane of the tunable metalens with a custom microscope positioned on a motion-controlled stage. We illuminated the whole aperture of the lens by passing a fiber-coupled and collimated 1550 nm SLD through a beam expander and adjusted the displacement of each plate by means of separate translation stages, projecting the microscope image of the focal plane onto an InGaAs camera (see supplementary Figure S2A). To characterize the focal length of the 633 nm lens, we illuminated the device with a HeNe laser passed through a beam expander, actuated the plates with micrometer translations stages, and measured the focal length with a meterstick (see supplementary Figure S2B).

Imaging with the metalens:

To image with our device, we illuminated our test patterns with a 625 nm LED, placed the tunable lens in the optical path at the fixed object distance, and tuned the focal length with translation stages to project the image directly onto a CCD camera (see supplementary Figure S3). To zoom, we tuned the focal length of our metalens and shifted the CCD camera to the image plane. We imaged both by illuminating patterns in transmission and by scattering light off printed object patterns.

Efficiency Measurements:

To measure the focusing efficiency of the lenses, we took the ratio of the optical power measured with a power meter (Newport Model 843-R) at the focal plane to that measured near the surface of the lens after transmission. In the case of the 633 nm efficiency measurement with the quartz substrate design, due to the super-wavelength lattice constant and higher diffraction orders which we could not capture with our power meter, we defined our focusing efficiency as the ratio of the focal plane power to that which passes through two glass plates with quartz substrates mounted on each with the same orientation and separation distance as our Alvarez lens setup. In using these glass plates and quartz substrates without any nitride patterning, this provides an underestimate of the focusing efficiency as the patterned nitride layer would induce further reflections and scattering, reducing the total optical power transmitted through the lens surface. The zeroth order beam efficiency of the metalens is defined as the ratio of the focal plane power to that measured passing through the metalens surface itself with the higher diffraction orders not being captured by the power meter.

Code Availability:

A version of the custom Python script used for writing layout files is included in the supplementary materials.

**Acknowledgements:**

This work was facilitated though the use of advanced computational, storage, and networking infrastructure provided by the Hyak supercomputer system at the University of Washington (UW). Part of this work was conducted at the Washington Nanofabrication Facility / Molecular Analysis Facility, a National Nanotechnology Coordinated Infrastructure (NNCI) site at the University of Washington, which is supported in part by funds from the Molecular Engineering & Sciences Institute, the Clean Energy Institute, the Washington Research Foundation, the M. J. Murdock Charitable Trust, the National Science Foundation and the National Institutes of Health. The research work is supported by an Amazon Catalyst Award and a SAMSUNG-GRO grant.

**Author Contributions:**

A.M., A.Z., and S.C. conceived the idea. S.C. conducted the simulations and design, fabricated the devices, and characterized them. S.C. wrote the paper with feedback from everyone else. A.M. supervised the whole project.

**Competing Interests:**

The authors declare no competing interests.

**Figure Legends:**

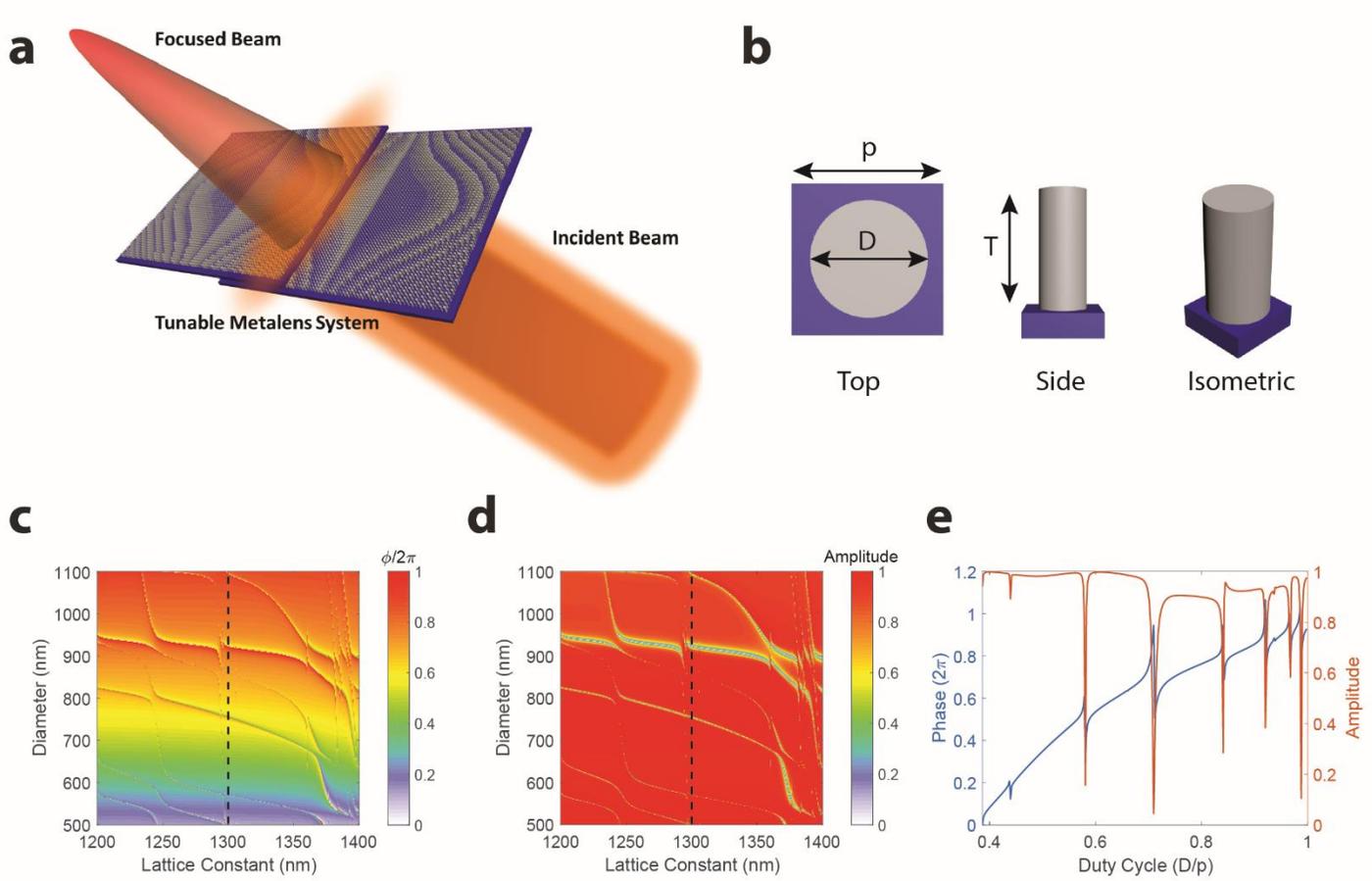

**Figure 1: Simulation and Design of Nanoposts** (a) Schematic representation of our tunable metalens system comprising two cubic metasurface phase plates actuated laterally. (b) Top, side, and isometric views of our silicon nitride nanoposts where $T$ is thickness, $D$ is diameter, and $p$ is lattice constant. The simulated amplitude (c) and phase (d) of the transmission coefficient as a function of nanopost diameter and lattice constant are shown. (e) The phase and amplitude for a fixed lattice constant of 1.3 μm corresponding to the black dashed lines in (c) and (d).

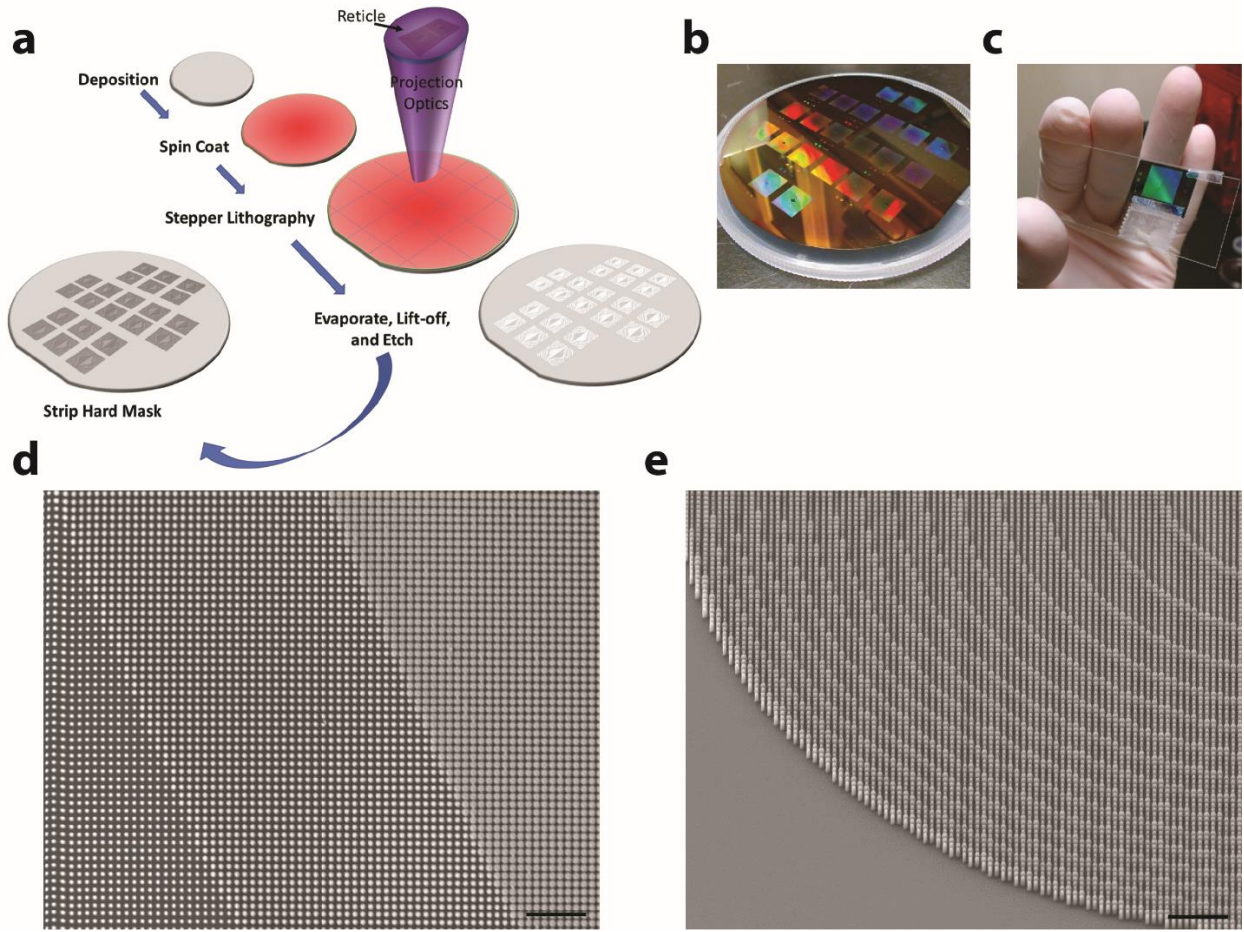

**Figure 2: Fabrication of the large area metasurfaces.** (a) Schematic of the process flow for fabricating multiple large area metalenses in parallel using high-throughput stepper lithography-based processing. (b) A fully exposed and developed 100 mm wafer, showing the capability to make large area devices. (c) A fully etched and cleaved metasurface cubic phase plate with a hand for scale. Scanning electron micrographs of fabricated nanoposts are shown at normal incidence (d) and 45° incidence (e). Scale bars 10 μm.

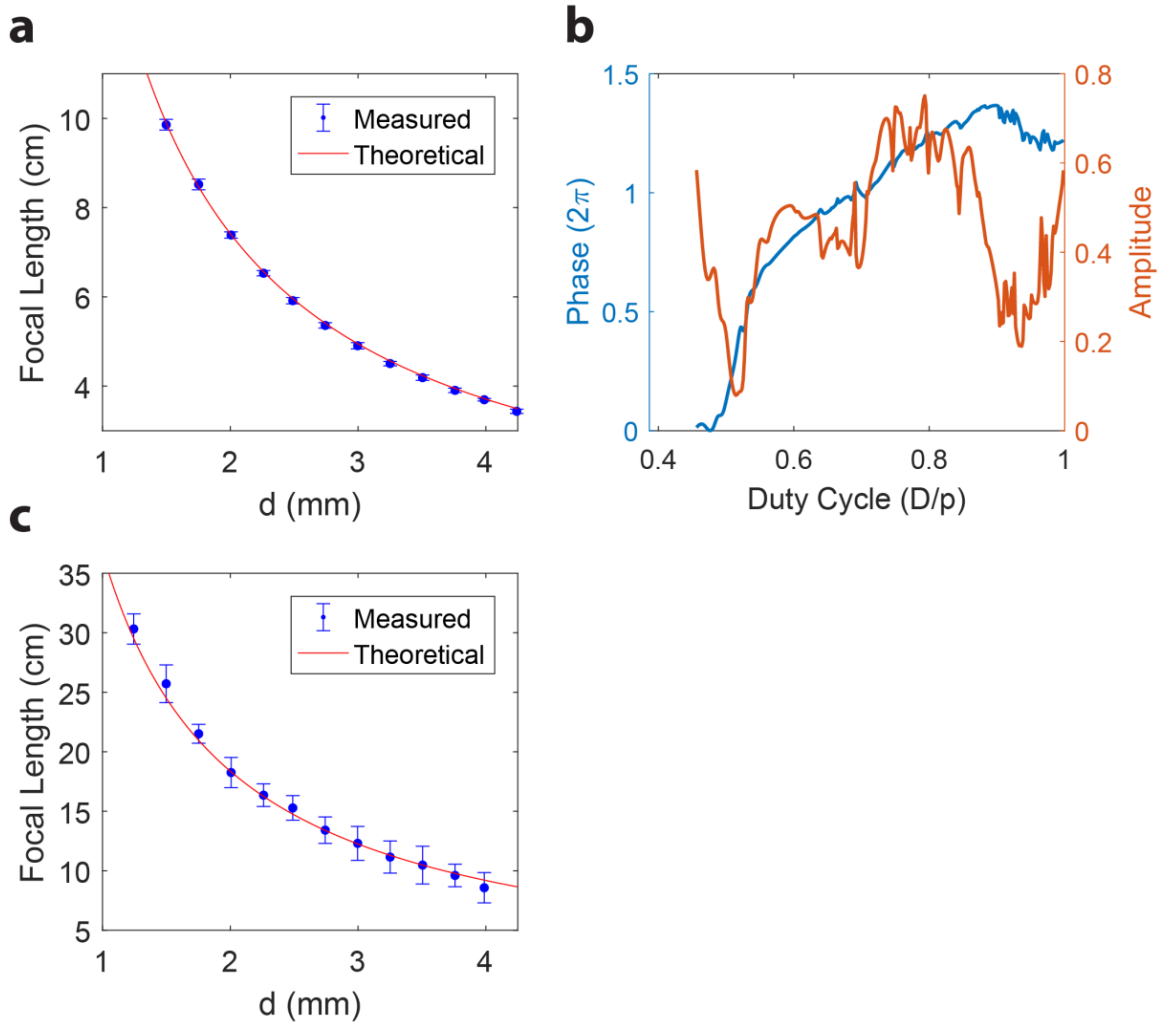

**Figure 3: Experimental and theoretical focal lengths of the tunable lens designs.** Focal length as a function of lateral displacement for the infrared (a) and visible (c) designs are shown. Errors bars represent a 95% confidence interval where the 1σ uncertainty is estimated during measurement by finding the range of distances over which the lens appears to be in focus. (b) Simulated transmission coefficient of the 1.5 μm thick silicon nitride nanoposts on a quartz substrate.

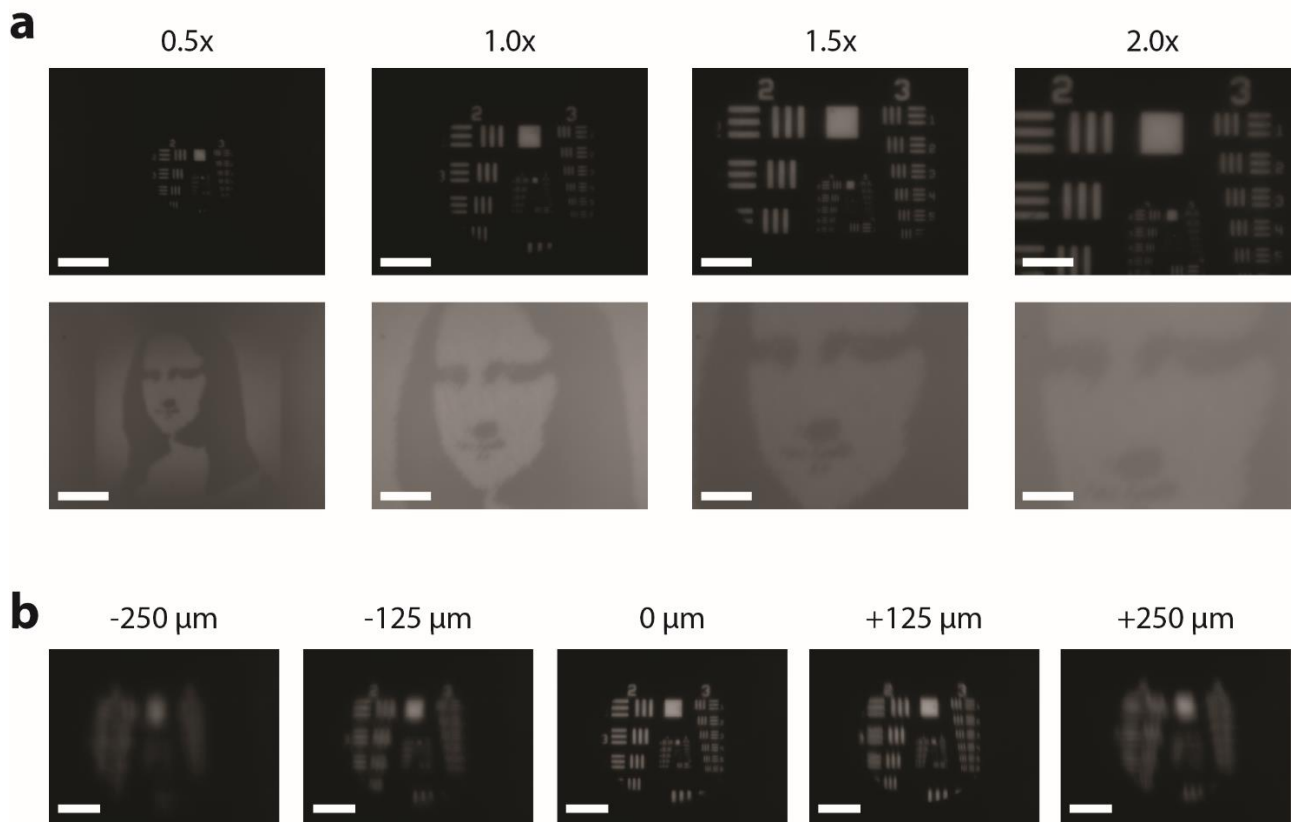

**Figure 4: Imaging with different magnifications using the visible lens design.** (a) Examples of images captured without any optical elements aside from the tunable lens system in the optical path using an Air Force resolution test chart illuminated in transmission (top row) and a Mona Lisa pattern printed on paper illuminated by scattering light off the pattern (bottom row). (b) The effect of misaligning the regular cubic metasurface phase plate on image quality. The scale bar is 1.2 mm for all images.

# Varifocal zoom imaging with large area focal length adjustable metalenses
## SUPPLEMENTARY INFORMATION

Shane Colburn[1], Alan Zhan[2], Arka Majumdar[1,2]*

[1]Department of Electrical Engineering, University of Washington, Seattle, Washington 98195, USA.

[2]Department of Physics, University of Washington, Seattle, Washington 98195, USA.

*Correspondence to: arka@uw.edu

**Characterization of lenses and vortex beam generators based on the silicon nitride nanopost platform**

In addition to the designed metasurface Alvarez lenses of the main text, to further demonstrate the versatility of our nanopost design and fabrication process, our reticle also included several static singlet and vortex beam-generating metalenses for 1550 nm operation. Figure S1 summarizes the data collected in characterizing these devices, demonstrating successful focusing at the design focal length (Figure S1A-B) and orbital angular momentum generation for different values of topological charge (Figure S1C). Our designed 2 mm static focal length lens exhibited close to diffraction-limited performance with a spot size of 3.94 µm, near the diffraction-limited value of 3.2 µm. The success of these devices verified the behavior and fabrication of the designed dielectric scatterers using our stepper lithography-based platform.

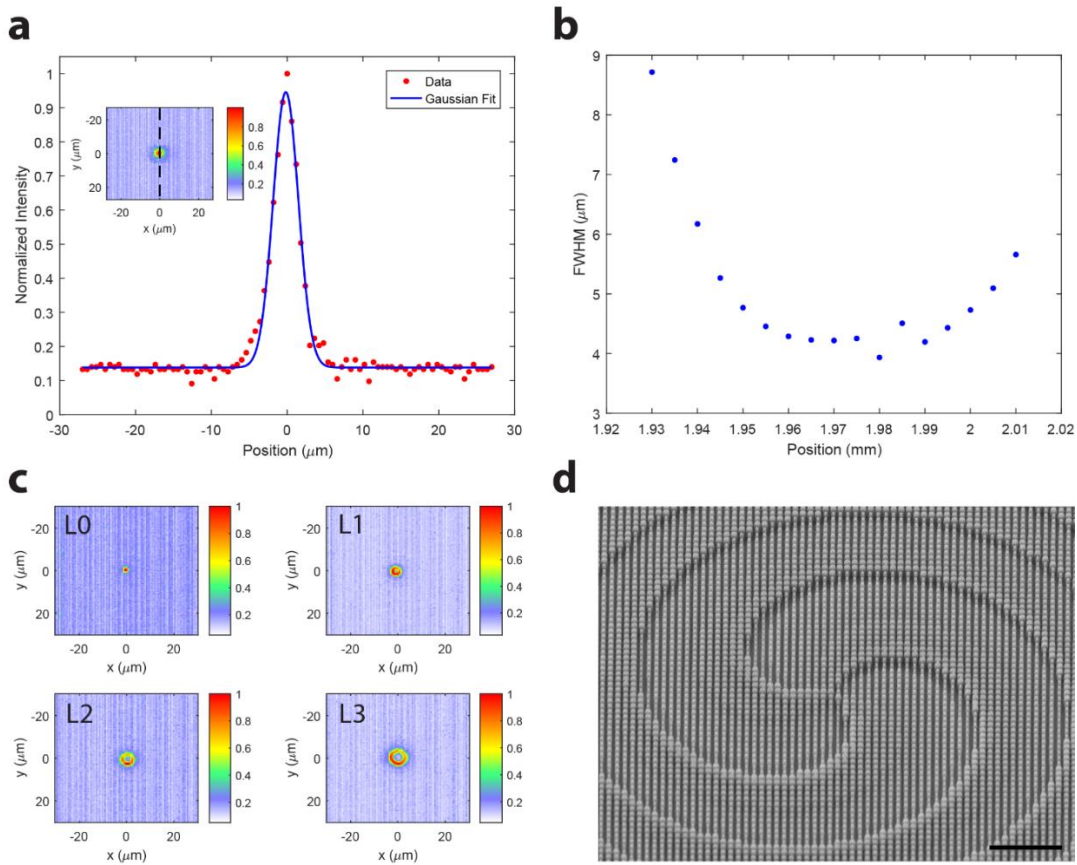

**Fig. S1: Characterization of static metasurface aspherical and vortex beam-generating lenses for 1550 nm wavelength.** (a) Cross section of a measured intensity profile at the focal plane of a designed static 2 mm focal length lens with 1 mm diameter using our nanoposts. The inset shows the 2D profile where the black dashed line indicates the position of the cross section. (b) Beam spot size in terms of full width at half maximum as a function of position along the optical axis of our lens from (a). (c) Measured cross sections at the focal plane of vortex beam-generating lenses, showing doughnut-shaped beams with different labelled values of topological charge. (d) 45° incident scanning electron micrograph of the L3 vortex beam lens. Scale bar 10 µm.

## Setup for measuring the focal length of the tunable metalens system

To measure the focal length of our metalens system, we had to illuminate our device with light passed through a beam expander (Thorlabs BE10M-A) to illuminate the full aperture of the metalens. In the case of the infrared design, we used a microscope on a motioned-controlled stage to image the focal plane of the metalens system (Fig. S2A). This microscope comprised an objective lens (Nikon Plan Fluor 20x/0.50, DIC M, ∞/0.17, WD 2.1), tube lens (Thorlabs ITL200), and camera (Xenics Bobcat-1.7-320). By laterally actuating the metasurface plates by means of translation stages, the focal length of the system would shift and we could translate the microscope until it was in focus and use the displacement to calculate the focal length. For the visible design, with the larger focal range supported by our device, it was beyond the range of our motion controlled-stage and microscope, so we used a reflective screen and meterstick to measure the focal length (Fig. S2B). For the infrared design, we used a 1550 nm SLD (Thorlabs S5FC1005P) and for the visible design we used a HeNe laser as our illumination sources.

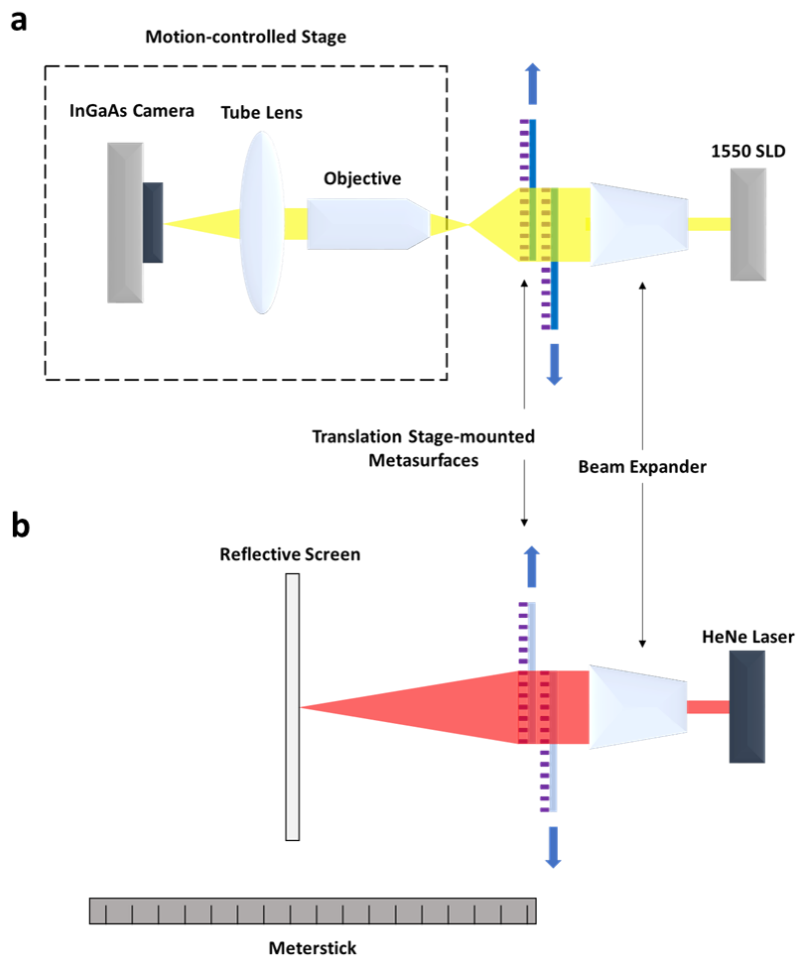

**Fig. S2: Experimental setups for measuring the focal length of our tunable system.** (a) Setup for measuring the focal length of our infrared tunable metalens. (b) Schematic of experimental setup for measuring the focal length of our visible wavelength tunable lens design.

## Setup for imaging in transmission and scattering modes

To image patterns with our metalens system, we illuminated objects in transmission (Fig. S3A) and by scattering light off printed object patterns with off-axis illumination (Fig. S3B) using a 625 nm LED (Thorlabs M625F2). Without additional optical elements in the path, we adjust the focal length of our metalens with translation stages to project an image of the object pattern directly onto a camera (AmScope MU300).

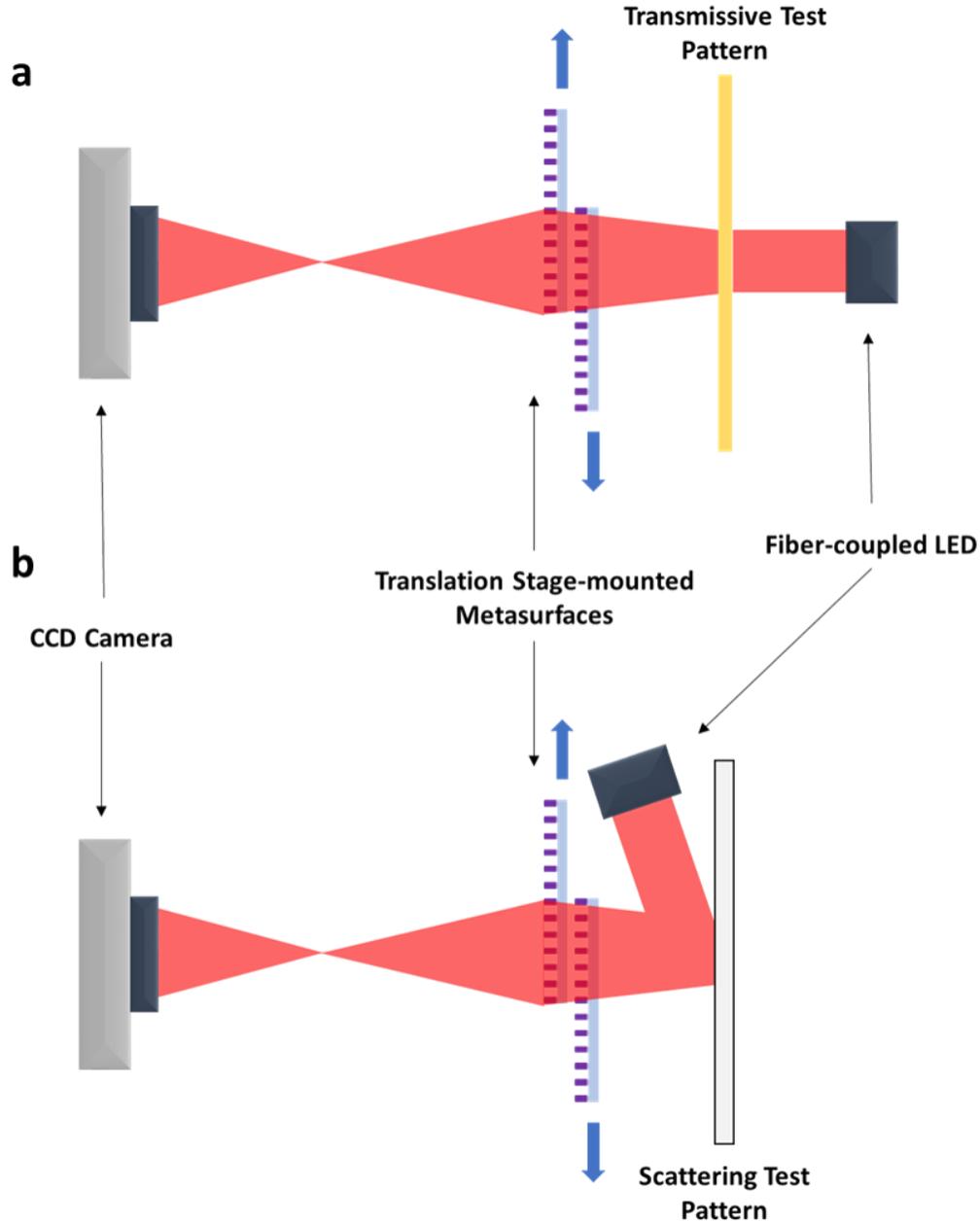

**Fig. S3: Experimental setup for imaging experiments.** We imaged object patterns in transmission (a) and by scattering light off objects (b) by projecting the image directly onto a camera without any additional magnification optics.

**Layout file generation algorithm**

In our algorithm, we use a dictionary data structure containing 6 unique nanopost radii as keys, each of which maps to a unique 32-sided regular polygon cell which approximates a circle. The 6 possible key-value pairings correspond to 6 possible discrete phase shifts in the 0 to $2\pi$ range to reduce the required memory, as opposed to utilizing a continuous spectrum of diameters. In generating the file, we iterate through a list of diameters and positions describing our layout and instantiate the appropriate cell reference for each nanopost by accessing our dictionary with the nanopost's radius key, as opposed to making a separate cell for each nanopost. In the case of our metasurface Alvarez lens metasurface plates, we cut the memory from 372 GB to 4.4 GB as a GDSII, and to only 140 MB as an OASIS file (more than a 2600x reduction).